\documentclass[draft,jgrga]{agutex2015}

  \usepackage{graphicx}
  \usepackage{color}
  \usepackage{amssymb}
  \usepackage{caption}
  \usepackage{rotating}

\authorrunninghead{GENESTRETI ET AL.}

\titlerunninghead{$\vec{J}\cdot\vec{E}'$ IN COMPONENT RECONNECTION}

\begin{document}

%
%

\title{The effect of a guide field on local energy conversion during asymmetric magnetic reconnection: MMS observations}
%
%

%
%

 \authors{K. J. Genestreti,\altaffilmark{1}
 J. L. Burch,\altaffilmark{2} 
 P. A. Cassak, \altaffilmark{3} 
 R. B. Torbert,\altaffilmark{4,2}
 R. E. Ergun, \altaffilmark{5,6} 
 A. Varsani, \altaffilmark{1}
 T. D. Phan, \altaffilmark{7} 
 B. L. Giles, \altaffilmark{8}
 C. T. Russell, \altaffilmark{9} 
 S. Wang, \altaffilmark{10}
 M. Akhavan-Tafti, \altaffilmark{11} 
 R. C. Allen \altaffilmark{12,2} 
 }

\altaffiltext{1}{Space Research Institute, Austrian Academy of Sciences, Graz, Austria.}
\altaffiltext{2}{Space Science and Engineering Division, Southwest Research Institute, San Antonio, TX, USA.}
\altaffiltext{3}{West Virginia University, Morgantown, WV, USA.}
\altaffiltext{4}{Space Science Center, University of New Hampshire, Durham, NH, USA.}
\altaffiltext{5}{Department of Astrophysical and Planetary Sciences, University of Colorado, Boulder, CO, USA.}
\altaffiltext{6}{Laboratory for Atmospheric and Space Physics, University of Colorado, Boulder, CO, USA.}
\altaffiltext{7}{Space Science Institute, University of California Berkeley, Berkeley, CA, USA.}
\altaffiltext{8}{Heliophysics Science Division, NASA Goddard Space Flight Center, Greenbelt, MD, USA.}
\altaffiltext{9}{University of California, Los Angeles, CA, USA}
\altaffiltext{10}{University of Maryland, College Park, MD, USA}
\altaffiltext{11}{Climate and Space Sciences and Engineering Department, University of Michigan, Ann Arbor, MI, USA.}
\altaffiltext{12}{Department of Physics and Astronomy, University of Texas San Antonio, San Antonio, TX, USA.}









%
%


\keypoints{\item{Determined location where $\vec{J}\cdot\vec{E}'>0$ for 11 asymmetric EDRs with different guide fields.}
		\item{Increasing guide field strength appears to move $\vec{J}\cdot\vec{E}'>0$ from electron-crescent to X-point.}
		\item{Guide field allows electron streaming at X-point, which takes work by the electric field.}}


%
%


\begin{abstract}
We compare case studies of Magnetospheric Multiscale (MMS)-observed magnetopause electron diffusion regions (EDRs) to determine how the rate of work done by the electric field, $\vec{J}\cdot(\vec{E}+\vec{v}_e\times\vec{B})\equiv\vec{J}\cdot\vec{E}'$ varies with shear angle. We analyze MMS-observed EDR event with a guide field approximately the same size as the magnetosheath reconnecting field, which occurred on 8 December 2015. We find that $\vec{J}\cdot\vec{E}'$ was largest and positive near the magnetic field reversal point, though patchy lower-amplitude $\vec{J}\cdot\vec{E}'$ also occurred on the magnetosphere-side EDR near the electron-crescent point. The current associated with the large $\vec{J}\cdot\vec{E}'$ near the X-point was carried by electrons with a velocity distribution function (VDF) resembling the magnetosheath inflow, shifted in the $-v_\parallel$ direction. At the magnetosphere-side EDR, the current was carried by electrons with a crescent-like VDF. We compare this 8 December event to 10 other EDRs with different guide field strengths. The dual-region $\vec{J}\cdot\vec{E}'$ was observed in three other moderate-shear EDR events, whereas three high-shear events had a strong positive $\vec{J}\cdot\vec{E}'$ near the electron-crescent point and one low-shear event had a strong positive $\vec{J}\cdot\vec{E}'$ only near the $B_L=0$ point. The dual-region $\vec{J}\cdot\vec{E}'>0$ was seen for one of three ``intermediate"-shear EDRs with guide fields of $\sim$0.2--0.3. We propose a physical relationship between the shear angle and mode of energy conversion where (a) a guide field provides an efficient mechanism for carrying a current at the field reversal point (streaming) and (b) a guide field may limit the formation of crescent eVDFs, limiting the current carried near the stagnation point.
\end{abstract}

%
%

%

\begin{article}

%
%

\section{Introduction}

Magnetic reconnection is a fundamental process in plasmas. It is a change in the topology of a magnetized plasma boundary coupled with the exchange of energy from magnetic fields to particles. The topological change occurs in the electron diffusion region (EDR), wherein the electrons are demagnetized, i.e., $\vec{E}+\vec{v}_e\times\vec{B}\neq0$. The per-volume rate of work done by the electric field on the plasma, which is often expressed in the electron rest frame as $\vec{J}\cdot(\vec{E}+\vec{v}_e\times\vec{B})\equiv\vec{J}\cdot\vec{E}'$ \citep{Zenitani.2011}, occurs in a region sometimes called the `dissipation region' in order to distinguish it from the EDR \citep{PritchettandMozer.2009}. $\vec{J}\cdot\vec{E}'$ specifically represents the rate of work done on the plasma by non-ideal electric fields. Because the $\vec{E}+\vec{v}_e\times\vec{B}\neq0$ is a defining condition for both the EDR and the $\vec{J}\cdot\vec{E}'$ region, the two regions may partially overlap; however, observations \citep{Burch.2016b} and simulations \citep{Zenitani.2011} show that significant $\vec{J}\cdot\vec{E}'$ may occur several electron inertial lengths away from the magnetic X-point, where the magnetic topology changes.

Reconnection at the low-latitude magnetopause of Earth is typically asymmetric, as the plasma density in the magnetosheath can exceed the magnetospheric plasma density by an order of magnitude \citep{PhanandPaschmann.1996}. This density asymmetry alters the momentum balance equation in the vicinity of the EDR and causes the electron flow stagnation point, where there is no bulk electron motion, to be displaced from the X-point, where the magnetic field in the reconnection plane is a minimum \citep{CassakandShay.2007}. Guide field or component reconnection occurs when the local shear angle between the magnetosheath and magnetospheric magnetic fields is less than 180$^\circ$. The presence of a guide field causes the magnetic field strength at the X-point to be non-zero, which can magnetize electrons near the X-point and reduce the size of the electron gyroradius relative to the size of the current layer \citep{Swisdak.2005}.

Observations of asymmetric and nearly anti-parallel reconnection have showed that field-to-plasma energy conversion and parallel electron heating occur Earthward of the X-point \citep{Burch.2016b,Hwang.2017}. In the central (asymmetric and anti-parallel) EDR, the current associated with $\vec{J}\cdot\vec{E}'$ is carried by electrons with broad crescent-shaped velocity distribution functions (VDFs) that separate in the parallel and perpendicular directions near the stagnation point \citep{Burch.2016b,Hesse.2014,Shay.2016}. In the outer EDR, where the Hall magnetic field is observed but electron kinetic motion still allows for non-zero $\vec{E}'$, parallel crescents carry the current associated with $\vec{J}\cdot\vec{E}'$ \citep{Shay.2016,Hwang.2017}. The electrons may be demagnetized \citep{PritchettandMozer.2009,Hesse.2014,Burch.2016b,Hwang.2017} at the X-point, but the out-of-plane current there, a result of electron cusp motion \citep{Shay.2016}, is generally weak.

Little work has been done to determine how and why the location of the $\vec{J}\cdot\vec{E}'$ region may change with the magnetic shear angle. \citet{PritchettandMozer.2009} compared particle-in-cell simulations of reconnection with $B_M=0$ and $B_M=B_{L,sh}$ and found $J_\parallel E_\parallel$ at the X-point to be larger for the guide field case, though a physical explanation for this difference was not discussed. \citet{Hesse.2016} showed that electron-crescent VDFs also appeared near the electron stagnation point in a simulation of guide field ($B_M\sim B_{L,sh}$) reconnection. The intensity of the crescent relative to the core of the VDF was significantly reduced in intensity as compared to their similar simulation of anti-parallel reconnection \citep{Hesse.2014}. According to \citet{Hesse.2016}, crescent VDFs should reduce in intensity and eventually disappear as the guide field intensity increases to the point where the magnetic scale length $B_L/(\partial B_L/\partial N)$ exceeds the electron Larmor radius, preventing mixing of electrons by thermal motion between regions with significantly different magnetic field directions.

In this study, we analyze the 8 December 2015 (11:20 UT) EDR event of \citet{BurchandPhan.2016}. We determine the electron-frame energy conversion rate and analyze the eVDFs associated with the current. We find that three of the four MMS spacecraft observed strong $\vec{J}\cdot\vec{E}'>0$ near the X-point associated with $E_\parallel$-accelerated magnetosheath-inflow electrons. The eVDF is structured, with a higher-energy beam-like portion anti-aligned with the parallel electric field and a low-energy crescent-like portion. All of the four spacecraft also observed smaller positive $\vec{J}\cdot\vec{E}'$ Earthward of the X-point and strong negative $\vec{J}\cdot\vec{E}'$ where the high-energy beam-like portion of the eVDF wraps from the parallel direction into the perpendicular ($\vec{E}\times\vec{B}$) direction. 

The location of the $\vec{J}\cdot\vec{E}'>0$ region for this event is then compared with 10 other EDR events with different guide field strengths. For three high-shear ($B_M/B_{L,sh}\approx0$) events, the $\vec{J}\cdot\vec{E}'>0$ region was near the electron-crescent point, Earthward of the $B_L=0$ point. For one of three ``intermediate"-shear ($B_M/B_{L,sh}\approx0.2$) cases, $\vec{J}\cdot\vec{E}'>0$ was observed at both the $B_L=0$ and electron-crescent points, whereas $\vec{J}\cdot\vec{E}'>0$ only occurred at the electron-crescent points for the remaining two cases. For three moderate-shear ($0.5\leq B_M/B_{L,sh} \leq1$) events, $\vec{J}\cdot\vec{E}'>0$ was observed at both the $B_L=0$ and electron-crescent points, similar to the 8 December 2015 event. Lastly, for the single low-shear ($B_M/B_{L,sh} \approx 3$) case, $\vec{J}\cdot\vec{E}'>0$ was observed only at the $B_L=0$ point and no clear electron crescents were observed.

We suggest that, based on these observations, the strength of the guide field may be a crucial factor in determining where electric fields convert their energy during asymmetric reconnection. The absolute and relative locations of the X, electron stagnation, and maximum $\vec{J}\cdot\vec{E}'$ points depend on a number of additional factors (see our companion study of Cassak et al. [submitted; this Volume]), including but not limited to the degree of asymmetry between the upstream plasma number densities, the strengths of the reconnecting component of the magnetic field, and the ion and electron temperatures. The absolute distances between the X, electron stagnation, and maximum $\vec{J}\cdot\vec{E}'$ points will also depend on the distance from its center where the EDR is observed, as well as the path of the spacecraft through the EDR. We do not attempt to control for each of variables individually, as a full investigation of this parameter space is beyond the scope of this study. However, we note that similar features in $\vec{J}\cdot\vec{E}'$ were observed for the few ($\leq$4) events within each category of high, moderate, and low shear, despite significant differences in other upstream conditions.

In the following section we describe the MMS instrumentation and data analysis techniques used in this study. In section 3 we present our case analysis of the 8 December 2015 EDR event. In section 4 we compare and the 8 December 2015 EDR event against 10 others with differing guide field strengths and upstream parameters. In section 5 we summarize our findings and outline topics that warrant future investigation. Further discussion may also be found in our companion study, Cassak et al. [submitted; this Volume], which presents simulations of three of the events studied here, as well as a discussion of what may govern the $\vec{J}\cdot\vec{E}'>0$ location for 2-d steady-state reconnection.

\section{Instrumentation and data}

This study analyzes burst-mode data from the suite of plasma particle and field instruments onboard MMS \citep{Burch.2016a}. The fast plasma investigation (FPI) dual ion and electron spectrometers (DIS and DES, respectively) measure differential directional fluxes for their namesake particle species at 32 energies between $\sim$10 eV/$q$ and $\sim$28 keV/$q$ \citep{Pollock.2016}. FPI-DIS and DES measure 4$\pi$-steradian velocity distribution functions (VDFs) once every accumulation period, 150 ms for the ions and 30 ms for the electrons. The DC magnetic field vector is provided by the fluxgate magnetometers (FGM) at 128 samples per second \citep{Russell.2016}. The AC magnetic field vector is provided by the search coil magnetometers (SCM) at a rate of 8196 samples per second \citep{LeContel.2016}, as are the spin-plane \citep{Lindqvist.2016} and axial \citep{Ergun.2016} components of the electric field, which are measured by two sets of probes collectively referred to as the electric field double probes (EDP). All of the data used in this study are available through the MMS science data center (https://lasp.colorado.edu/mms/sdc/public/), with the exception of the level 3 (L3) EDP data used during the analysis of the 8 December event, which are available by request.

Some of the data are resampled and/or smoothed prior to analysis. We shifted all of the FPI data forward by half of an acquisition period (+0.075 seconds for DIS and +0.015 seconds for DES), such that the times associated with each data point mark the average, rather than the beginning, of the associated acquisition period. It is unnecessary to perform a similar shift for the fields data, since the time stamps are already centered on a significantly smaller measurement period. We have also smoothed the AC electric field data using a sliding overlapping boxcar scheme, where the width of the boxcar ($\pm$15 ms) was chosen to match the sample rate of FPI-DES and provide the best possible agreement between $\vec{E}$ and $-\vec{v}_e\times\vec{B}$. Smoothing the electric field reduces the magnitude of positive and negative oscillations of $\vec{J}\cdot\vec{E}'$, but makes the bulk action of the electric field on the plasma more easily discernible.

All data is shown in either magnetopause-normal (LMN) or field-aligned (FAC) coordinates. The LMN eigenvector system for the 8 December 2015 event is taken from \citet{BurchandPhan.2016}, which was determined using minimum variance and minimization of Faraday residue \citep{KhrabrovandSonnerup.1998}. The coordinate axes, $\hat{L}$, $\hat{M}$, and $\hat{N}$ are constant in time and defined in a GSE basis as [0.3641, --0.1870, 0.9124], [--0.2780, --0.9568, 0.08515], and [0.8889, --0.2226, --0.4003], respectively. $\hat{L}$ is the direction of maximum magnetic variance and the reconnection outflow. $\hat{M}$ is the direction of intermediate variance and the guide and Hall magnetic fields. $\hat{N}$ is the direction of minimum variance, the magnetopause normal, and the reconnection inflow. For FAC, the coordinate axes are calculated for each magnetic field measurement and are defined as $\hat{v}_\parallel$, $\hat{v}_{\bot1}$, and $\hat{v}_{\bot2}$, which are defined as $\hat{b}$, $(\hat{b}\times\hat{v}_e)\times\hat{b}$, and $\hat{b}\times\hat{v}_e$, respectively, where $\hat{b}$ is the direction of the magnetic field and $\hat{v}_e$ is the direction of the electron bulk velocity.

\section{Analysis of the 8 December 2015 (11:20 UT) EDR event}

\begin{figure*}
\noindent\includegraphics[width=35pc]{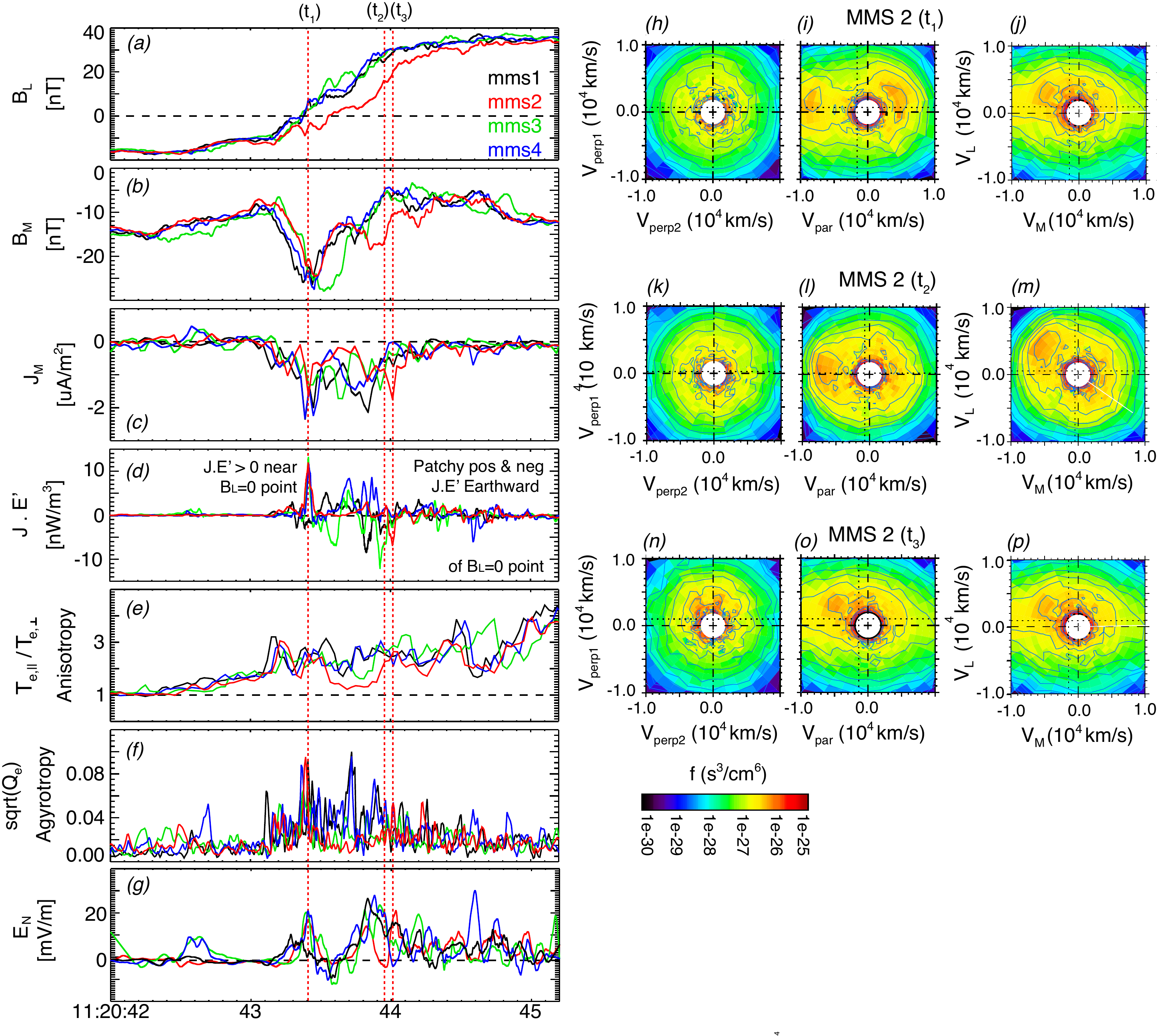}
\caption{Measurements from the four spacecraft, MMS1 (black), 2 (red), 3 (green) and 4 (blue). Panels are: (a) $B_L$, (b) $B_M$, (c) $J_M$, (d) the energy conversion rate $\vec{J}\cdot(\vec{E}+\vec{v}_e\times\vec{B})\equiv\vec{J}\cdot\vec{E}'$, (e) the electron anisotropy, defined as the ratio of the parallel and perpendicular temperatures, (f) the electron agyrotropy, defined by \citet{Swisdak.2016} with the $\sqrt{Q_e}$ parameter, and (g) the normal component of the electric field. (h--j) eVDF cuts measured at $t_1$, indicated by the first vertical dashed line drawn on (a--g). (k--m) eVDF cuts measured at $t_2$ and (n--p) eVDFs at $t_3$. eVDFs are taken from MMS2. All spacecraft data are shifted such that the first large positive peaks of $\vec{J}\cdot\vec{E}'$ are aligned. The data from MMS2, 3, and, 4 were shifted by +22 ms, --166 ms, and --16 ms, respectively (Note that this is unrelated to the shifting of the FPI data mentioned in Section 2). The LMN coordinate system is taken from \citet{BurchandPhan.2016}.}
\label{8Dec}
\end{figure*}

An overview of MMS data for the 8 December 2015 event is provided in \citet{BurchandPhan.2016}. As a brief review, \citeauthor{BurchandPhan.2016} identified a $\sim$2-second EDR encounter at 11:20:43--45 UT, which was observed by all four MMS spacecraft. The average spacecraft separation was 15 km or roughly 8 electron inertial lengths $d_{e,sh}$, given the upstream magnetosheath density of $\sim$8 cm$^{-3}$. Several seconds before the spacecraft passed through the EDR, effectively moving from the magnetosheath to the magnetosphere, the density in the upstream magnetosheath was approximately 2.5 times the density in the upstream magnetosphere. There was an asymptotic out-of-plane guide magnetic field $B_M$ approximately the same size as the magnetosheath reconnecting field $B_{L,sh}$. \citet{BurchandPhan.2016} noted that the out-of-plane current $J_M$ was bifurcated, peaking strongly near the $B_L=0$ point and several tens of $d_e$ Earthward of the $B_L=0$ point. This bifurcated current differed significantly from the antiparallel reconnection event of \citet{Burch.2016b}, on 16 October 2015, which only had a peak in $J_M$ at the electron-crescent point. Additionally, \citet{BurchandPhan.2016} found that the electrons were highly anisotropic, where the parallel temperature exceeded the perpendicular temperature, at both the $B_L=0$ and electron-crescent points. The antiparallel event of 16 October had parallel heating only at (and Earthward of) the electron-crescent point, where the electron-frame field-to-plasma energy conversion rate $\vec{J}\cdot\vec{E}'>0$ was similarly peaked. \citet{BurchandPhan.2016} did not calculate $\vec{J}\cdot\vec{E}'$ for the 8 December event, which is calculated here and shown in Figure \ref{8Dec}.

We have shifted the data from MMS2--4 in Figure \ref{8Dec} such that the first large $\vec{J}\cdot\vec{E}'>0$ peaks from the four spacecraft are artificially aligned in time. This organizes some reconnection-related data, primarily near the $B_L=0$ point, but does not organize all of the data. Vertical dashed lines $t_1$ and $t_3$ on Figure \ref{8Dec}(a--f) mark the two peaks of the bifurcated out-of-plane current $J_M$ as measured by MMS2. The separation of the two peaks are well resolved by the MMS data, as approximately 20 eVDF measurements are made between $t_1$ and $t_2-t_3$. Overall, the data in Figure \ref{8Dec} show that the $\vec{J}\cdot\vec{E}'$ region was highly structured and located primarily at and Earthward of the $B_L=0$ point. Three of the four spacecraft (MMS2--4) observed strong positive $\vec{J}\cdot\vec{E}'\approx10$ nW/m$^3$ near the $B_L=0$ point. For MMS2, this first $\vec{J}\cdot\vec{E}'>0$ peak was sunward of the $B_L=0$ point, while for MMS3 and 4 the first $\vec{J}\cdot\vec{E}'>0$ peak was Earthward of the $B_L=0$ point. At $t_1$, the electrons were strongly anisotropic and agyrotropic. In asymmetric reconnection, agyrotropy is expected when the Larmor motion of the electrons allows for mixing between the high and low-density inflow regions in the vicinity of the EDR. The large agyrotropy seen here indicates that the considerable guide field of $B_M/B_{L,sh}\approx1$ is not sufficient to fully magnetize the electrons and prevent this mixing. For comparison, the largest value of $\sqrt{Q_e}$ shown here, which has values ranging from 0 (fully gyrotropic) to 1 (fully agyrotropic), is approximately the same as its largest value for the nearly anti-parallel 16 October event of \citet{Burch.2016b}.

Patchy, lower-amplitude, and mostly positive $\vec{J}\cdot\vec{E}'$ was also observed by all four spacecraft between $t_1$ and $t_3$, several tens of electron inertial lengths Earthward of the $B_L=0$ point. (Given a magnetopause normal velocity of --44 km/s and a magnetosheath electron inertial length of $d_{e,sh}=1.9$ km \citep{BurchandPhan.2016}, the spacecraft effectively move $\sim$2.3 $d_e$ every 100 milliseconds). Large negative excursions of $\vec{J}\cdot\vec{E}'$ were also observed, though, like the patchy positive $\vec{J}\cdot\vec{E}'$, these peaks are not well-organized by the time shifting done to organize the $\vec{J}\cdot\vec{E}'>0$ peaks at $t_1$. The large temperature anisotropy, which was first observed near $t_1$, extends Earthward of the $B_L=0$ point, up to and beyond $t_3$. The strong electron agyrotropy occurs mostly between $t_1$ and $t_3$, as does the strong out-of-plane current $J_M$. As with the out-of-plane current, the normal electric field is bifurcated. For fully anti-parallel reconnection, $E_N$ is expected to have a small shoulder at the X-point and a much stronger peak Earthward of the X-point near the electron stagnation point \citep{Shay.2016,RWang.2017}.

Figure \ref{8Dec}(h--p) shows selected cuts of eVDFs measured by MMS2. Panels (h--j) show eVDF cuts taken at $t_1$, at the center of the first out-of-plane current peak near the $B_L=0$ point, where $\vec{J}\cdot\vec{E}'$ is strong positive. A more complete set of distribution functions was presented in \citet{BurchandPhan.2016}. Figure \ref{8Dec}(h--j) shows that the eVDF associated with the current near the $B_L=0$ point is highly structured, with a beam-like counter-streaming portion at higher energies that is partially balanced by a lower-energy crescent-like portion of the eVDF at lower energies. The parallel motion at this point is almost entirely in the out-of-plane direction due to the presence of the strong guide field. In the picture of \citet{Shay.2016}, the higher-energy portion of this eVDF should be meandering sheath electrons that have been already entered the EDR, been accelerated by the normal electric field, then meander back to the X-point. The lower-energy portion of the eVDF then should be the newly inflowing sheath electrons that have not yet been accelerated.  

Panels (n--p) show eVDFs at $t_3$, the second of the two out-of-plane current peaks, where $\vec{J}\cdot\vec{E}'$ is negative. At the second of the two current peaks, between $t_2$ and $t_3$, the high-energy beam-like portion of the eVDF persists (panels (k--m)), then wraps from the parallel direction into the $v_{\bot1}$ direction (panels (n--p)). The lower-energy portion of the eVDF does not appear, which is consistent with the idea that this lower-energy portion were sheath electrons that had not yet undergone acceleration by the large $E_N$. The wrapping of this beam-like portion of the eVDF from the parallel to perpendicular directions is similar to the wrapping of crescent-shaped eVDFs in high-shear EDRs \citep{Burch.2016b}. Several studies have found EDRs with $\vec{J}\cdot\vec{E}'<0$ \citep{Hwang.2017,RWang.2017}. This may occur as a result of time-dependent evolution, such as current sheet thinning \citep{RWang.2017}, or as a result of processes that may occur in a steady-state, such as the breaking of super-Alfvenic electron jets in the outer EDR. In the simplest terms, $\vec{J}\cdot\vec{E}'<0$ is a conversion of plasma energy to electromagnetic energy in the reference frame of the electrons. In our companion paper, this $\vec{J}\cdot\vec{E}'<0$ did not appear during a steady-state period of a 2.5-d particle-in-cell simulation of this event, which may imply that the $\vec{J}\cdot\vec{E}'<0$ here was either a result of time-dependent or 3-d processes. The exact cause (and effect) in this particular case, though, is beyond the scope of this current investigation.

In summary, the 8 December 2015 ($B_M/B_{L,sh}\sim1$) EDR event had the following characteristics:
 \begin{enumerate}
 \item{$\vec{J}\cdot\vec{E}'$ was strongly positive at or very near the $B_L=0$ point, where a strong out-of-plane current was carried by counter-streaming electrons moving against the local magnetic field direction ($\approx\hat{M}$,}
 \item{Patchy positive and negative $\vec{J}\cdot\vec{E}'$ Earthward of $B_L=0$, where the mostly anti-field-aligned, higher-energy portion of the eVDF wrapped from the parallel direction to the perpendicular direction}
 \item{The electrons were anisotropic over a broad region extending from the $B_L=0$ point to far Earthward of the $\vec{J}\cdot\vec{E}'$ region}
 \item{The electrons were agyrotropic over a narrow region, roughly coinciding with the $\vec{J}\cdot\vec{E}'$ region}
 \end{enumerate}

\section{Analysis of $\vec{J}\cdot\vec{E}'$ for EDRs with differing shears}

\subsection{Overview of event list}

Here we determine the energy conversion rate $\vec{J}\cdot\vec{E}'$ for 10 additional EDRs, all of which have been identified in previous studies. For many of these events, e.g., the high-shear EDR of \citet{Burch.2016b} and the low-shear EDR of \citet{Eriksson.2016}, the energy conversion rate, electron dynamics, and larger-scale context have already been studied extensively. For other events, including some of those identified by \citet{Fuselier.2017} and \citet{Wang.2017}, the energy conversion rate has not been calculated in any previous study to the knowledge of the authors. The set of events is presented in Table 1.

\begin{sidewaystable}
\caption{EDR events and upstream conditions, sorted into four categories based on the strength of the guide field.}
\centering
\begin{tabular}{| c || l | l | l | l | l | l || l | l |}
\hline
 & Date  & $\frac{B_M}{B_{L,sh}}$ & $\frac{B_{L,sh}}{B_{L,sp}}$ & $\frac{n_{sh}}{n_{sp}}$ & $\frac{T_{e,sh}}{T_{e,sp}}$ & $\frac{T_{i,sh}}{T_{i,sp}}$ & X,Y,Z [$R_E$] & Reference \\
\hline
\hline 
High    & 2015-10-16/13:07  &  0.1    &  0.6  &  16  &   0.2  & 0.2  &   8.3, 8.5, -0.7 & \citep{Burch.2016b} \\  
shear  & 2015-09-19/09:10  &  0.1    &  0.6   &  15  &  0.5  &  0.2  &  6.4, 7.7, 0.02 & \citep{Hwang.2017} \\
           & 2015-12-11/12:16  &  0.15   &  0.4  &  10  &   1    &  0.4  &  9.3, 1.8, -0.9  & \citep{Wang.2017} \\
\hline
\hline
``Intermediate" & 2015-12-06/23:38  & 0.2  & 1     &  40  &   0.07 & 0.2 &  8.5, -4.0, -0.6 & \citep{Khotyaintsev.2016} \\
shear               & 2015-12-08/00:06  & 0.2  & 0.4  &  20  &    0.3   & 0.4 &  9.0, -3.9, -0.6  & \citep{Graham.2017} \\ 
                        & 2016-01-10/09:13  & 0.3  & 0.5  &  6    &    0.7   & 0.8 &  8.8, -2.4, -0.8  & \citep{Fuselier.2017}  \\
\hline
\hline
Moderate  & 2016-11-28/07:36                     &  0.5  &  0.5  &   30   &  0.4  & 0.1  &   10.0, 3.1, -3.2  & \\
shear        & 2015-11-11/12:35                      &  1     &  0.8  &   12   &  1     & 0.7  &   6.6, -1.7, -0.1   & \citep{Wang.2017} \\
                 & 2015-12-08/11:20$^{(a)}$         &  1     &  0.4  &   2.5  &  0.5  & 0.4  &   10.2, 1.6, -1.0  & \citep{BurchandPhan.2016} \\ 
                 & 2015-12-14/01:17                     &  1     &  0.5  &   10   &   0.3  & 0.5  &  10.1, -4.3, -0.8 & \citep{Chen.2017} \\
\hline
\hline
Low shear  &  2015-09-08/11:01 &  5  &  0.5  & 2.5 &  0.3 & 0.3 &   4.9, 9.2, 0.1   & \citep{Eriksson.2016} \\
\hline
\end{tabular}
\tablenotetext{(a)}{Presented in section 3.}
\end{sidewaystable}

The EDR events listed in Table 1 were selected from a larger set of reconnection events on the following basis: first, the EDR must have been observed during a full crossing of the magnetopause, such that (a) the upstream conditions could be determined immediately before and after the crossing and (b) the energy conversion rate $\vec{J}\cdot\vec{E}'$ could be calculated at both the $B_L=0$ and electron-crescent points. Second, the path of the spacecraft through the magnetopause, judging by the profile of $B_L$, should be reasonably simple, e.g., we exclude events where the spacecraft passes through a portion of the EDR, doubles back, then crosses again. Lastly, we excluded events for which we were unable to obtain a stable LMN coordinate system in which the upstream conditions (namely $B_M$ and $B_L$) had some reasonably constant-in-time asymptotic value. For the most part, this last criterion is a repetition of our previous criteria, as it mostly excluded partial or complex crossings of the magnetopause. This list of events is far from exhaustive and is nearly entirely biased towards the first of the two MMS magnetopause phases due to the current (at the time of writing) availability of phase 1a surveys \citep{Fuselier.2017,Wang.2017}.

The events listed in Table 1 have a diverse set of upstream conditions, with the density asymmetry ranging from $\sim$2.5 to $\sim$40, magnetic field $B_L$ asymmetry ranging from $\sim$1 (no asymmetry) to $\sim$0.4, and asymmetric electron (ion) temperatures ranging from 1 to less than 0.1 (0.7 to 0.1). For this investigation, we are primarily concerned with the strength of the guide field relative to the reconnecting component of the magnetosheath field $B_M/B_{L,sh}$, and any potential impact of these additional parameters on the location of the $\vec{J}\cdot\vec{E}'$ is not controlled for (see Section 5).

As in Table 1, we separate these events into four categories; the three ``High-shear" events have $B_M/B_{L,sh}<0.2$, the three ``Intermediate-shear" events have $B_M/B_{L,sh}\approx0.2$, the four ``Moderate-shear" events have $B_M/B_{L,sh}\approx0.5-1$, and the single ``Low-shear" event has a $B_M$ $\sim5$ times larger than $B_{L,sh}$. The upstream conditions were determined several seconds before and after the EDR crossing, such that they are as close as possible to the upstream conditions during the EDR crossing.

\subsection{High-shear events}

\begin{figure*}
\noindent\includegraphics[width=43pc]{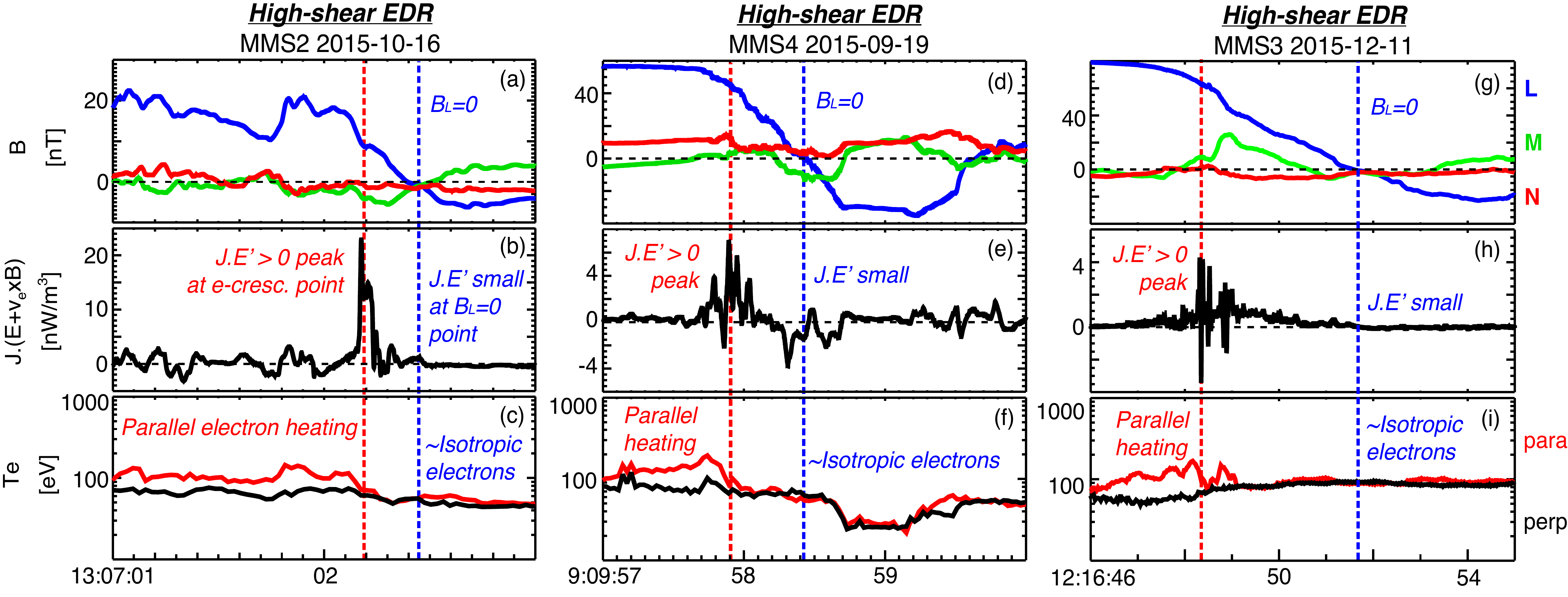}
\caption{(a, d, g) The magnetic field vector in LMN, (b, e, h) the local energy conversion rate, $\vec{J}\cdot\vec{E}'$, (c, f, i) the parallel (red) and perpendicular (black) electron temperatures, and (j, k, l) eVDFs at the energy conversion site (red outline) and at the X-point (blue outline). The first column shows the event of \citet{Burch.2016b}, the second column is the event of \citet{Hwang.2017}, and the third column \citet{Wang.2017}. For these three high-shear EDR events, which have guide fields approximately 10\% as large as the reconnecting sheath field, the maximum energy conversion rate is seen at the point where the electrons form crescent-shaped VDFs (red dashed line), not at the $B_L=0$ point (blue line). The LMN coordinate systems for the 2015-10-16 and 2015-09-19 events were taken from \citet{Burch.2016b} and \citet{Hwang.2017}, respectively. The LMN coordinate system for the 2015-12-11 event was determined by performing minimum variance analysis of the magnetic field vector (MVAB) measured by MMS3 between 12:16:38 -- 12:16:59 UT. In X, Y, and Z GSE, the axes are L = [0.408, --0.333, 0.850], M = [0.196, --0.87735372, --0.43783073], and N = [0.891, 0.346, --0.293].}
\label{hsplots}
\end{figure*}

Figure \ref{hsplots} shows $\vec{J}\cdot\vec{E}'$ for the three high-shear EDR events of (Fig \ref{hsplots}(a)--(c)) \citet{Burch.2016b}, (Fig \ref{hsplots}(d)--(e)) \citet{Hwang.2017}, and \citet{Wang.2017}, all of which have guide fields approximately 10--15\% as large as $B_{L,sh}$. The vertical dashed blue line marks the $B_L=0$ point where, for all three cases, there is no significant $\vec{J}\cdot\vec{E}'>0$. For the 19 September event, there is some $\vec{J}\cdot\vec{E}'<0$ at/near the field reversal point, which is discussed in \citet{Hwang.2017}. In all three cases, both the $\vec{J}\cdot\vec{E}'>0$ peaks and the parallel heating of electrons occurred on the magnetospheric side of the $B_L=0$ point at the electron-crescent point, which is marked in Figure \ref{hsplots} with a vertical dashed red line. In the case of the 16 October event, which is thought to have been an observation very near the center of an EDR, both perpendicular (Fig \ref{hsplots}j) and parallel crescents (Fig \ref{hsplots}k) were observed Earthward of the $B_L=0$ point, while the electrons at the $B_L=0$ point were largely isotropic and gyrotropic. The 19 September and 11 December cases were likely outside the central EDR, as only parallel crescents were observed. This is consistent with the Hall deflections of $B_M$ that were seen during these two EDR encounters, which indicated that the spacecraft passed some distance along $L$ from from the point of symmetry.

The separation between the $\vec{J}\cdot\vec{E}'$ peaks and the $B_L=0$ points were well-resolved for all three of these events due to the very high time resolution of MMS measurements. For the 16 October event, $\sim10$ eVDFs were collected between the $B_L=0$ point and the $\vec{J}\cdot\vec{E}'$ peak; $\sim20$ eVDFs were collected between these points for the 19 September event, and $\sim$110 eVDFs were collected between these points for the 11 December event.

\subsection{Moderate-shear events}

\begin{figure*}
\noindent\includegraphics[width=43pc]{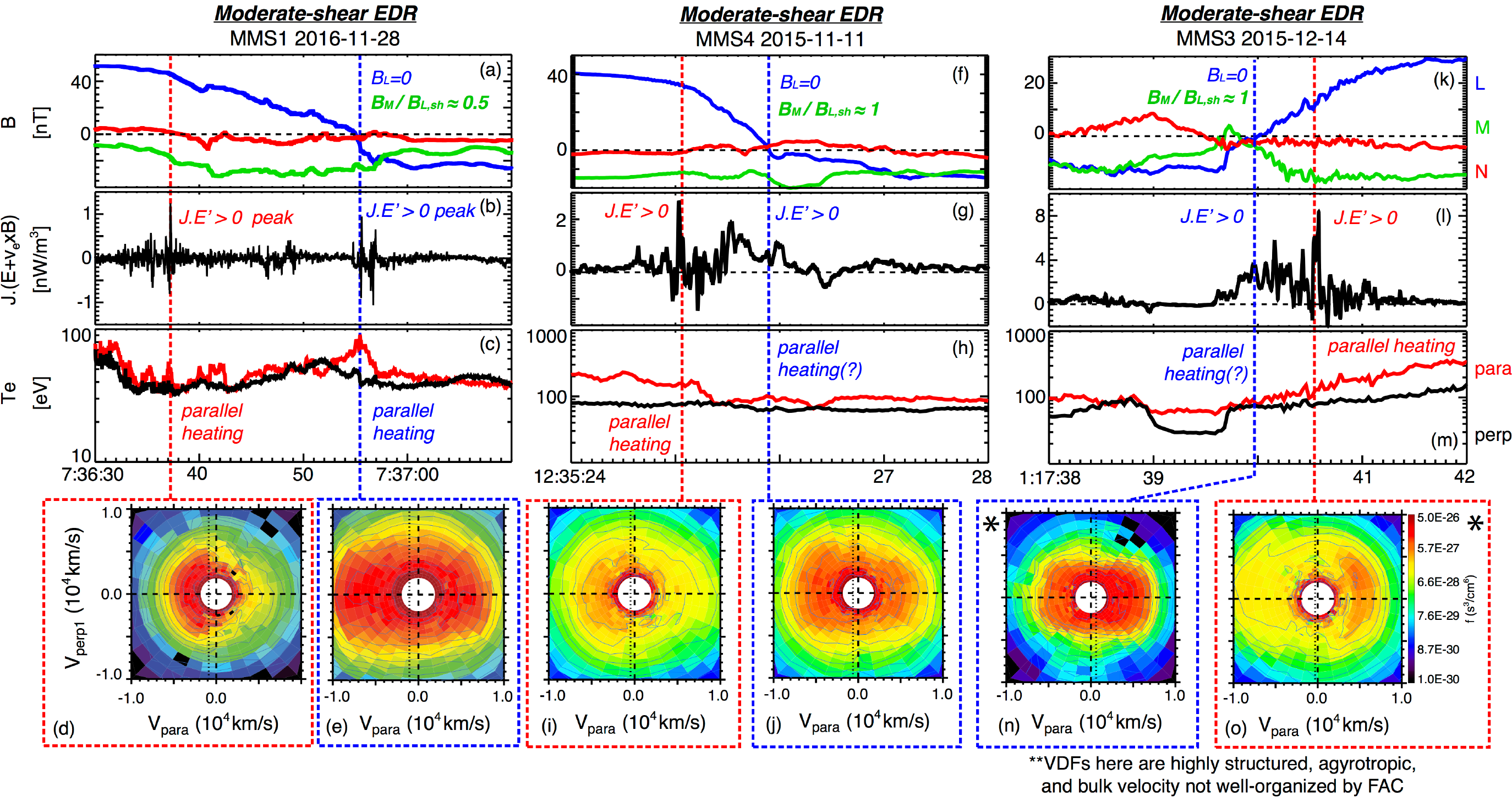}
\caption{(a, f, k) The magnetic field vector in LMN, (b, g, l) the energy conversion rate, $\vec{J}\cdot\vec{E}'$, and (c, h, m) the parallel (red) and perpendicular (black) electron temperatures for three moderate-shear events. (d, i, o) eVDF cuts taken at the electron-crescent and (e, j, n) at the $B_L=0$ points. eVDFs are shown in the $v_{\bot1}-v_{\parallel}$ plane. The first column shows the previously unidentified 2016-11-28 event, the second column shows data from an event of \citet{Wang.2017}, and the third column shows data from the event of \citet{Chen.2017}. The dashed lines mark the $B_L=0$ (blue) and electron-crescent (red) points. The LMN system for the 2016-11-28 event was determined by applying the minimization of Faraday residue technique to MMS1 data measured between 7:36:32 -- 7:36:50 UT. In X, Y, and Z GSE, the axes for this event are L = [0.178, --0.159,  0.971], M = [0.245, --0.949, --0.200], and N = [0.953,  0.273, --0.130]. The LMN system for the 2015-11-11 event were determined with MVAB of the data from MMS4 between 12:35:21 -- 12:35:29 UT, which yielded L = [0.376, --0.0458, 0.925], M = [--0.448, --0.883, 0.138], and N = [0.811, --0.466, --0.353].}
\label{msplots}
\end{figure*}

Figure \ref{msplots} shows $\vec{J}\cdot\vec{E}'$ for three moderate-shear EDR events, which had guide fields 50--100\% as large as the magnetosheath $B_L$. The event shown in the middle column of Figure \ref{msplots} was identified by \citet{Wang.2017} and the event shown in the right-most column was studied by \citet{Chen.2017}. The left-most column has not been identified yet as an EDR to the knowledge of the authors. 

The locations of the $\vec{J}\cdot\vec{E}'>0$ peaks for these three events are qualitatively similar to that of the 8 December (11:20 UT) event, in that $\vec{J}\cdot\vec{E}'>0$ and the parallel heating of electrons occur at both the $B_L=0$ and electron-crescent points. For all three cases, the current Earthward of the $B_L=0$ point was carried by electrons with parallel crescent-shaped VDFs. In the case of the 14 December event, highly agyrotropic perpendicular crescent eVDFs were also observed. For the 28 and 11 November and events, the current at the $B_L=0$ was carried by electrons with VDFs similar to the anisotropic magnetosheath inflow, but shifted in the local $-v_\parallel$ direction, against the guide field in $+M$. For the 14 December event, the current at the $B_L=0$ point had parallel and perpendicular components and the VDFs were not clearly organized by the local magnetic field coordinates. In all three cases, the electrons are broadly anisotropic around both the $B_L=0$ and crescent points, though for the 11 November and 14 December cases, it is difficult to determine if this anisotropy is a result of local heating or is an extension of the anisotropy generated in the upstream magnetosheath inflow region.

For the 11 November case, as for the 8 December (11:20 UT) case, the crescent-shaped portion of the eVDF is not as intense, relative to the background plasma, as it was for the very high-shear events. This is not the case for the $B_M/B_{L,sh}\sim0.5$ event of 28 November or for the $B_M/B_{L,sh}\sim1$ event of 14 December, both of which had pronounced crescent-shaped eVDFs. The electron agyrotropy, as defined by the Swisdak parameter $\sqrt{Q_e}$ \citep{Swisdak.2016}, was nearly twice as large at the $B_L=0$ point than at the electron-crescent point for all three $B_M/B_{L,sh}=1$ events (including the 8 December (11:20 UT) event). For the 28 November event, which had $B_M/B_{L,sh}=0.5$, the agyrotropy was equally as strong at both points. For the high-shear events, the agyrotropy was sharply peaked at the electron-crescent points alone. There was no significant difference between the maximum values of $\sqrt{Q_e}$ for the high and moderate-shear EDR cases, as the differences between events in a given shear category were comparable to the differences between events in different categories. This may be due to the small sample size and the large spread in upstream parameters within each category.

\subsection{Low-shear event} 

\begin{figure*}
\noindent\includegraphics[width=43pc]{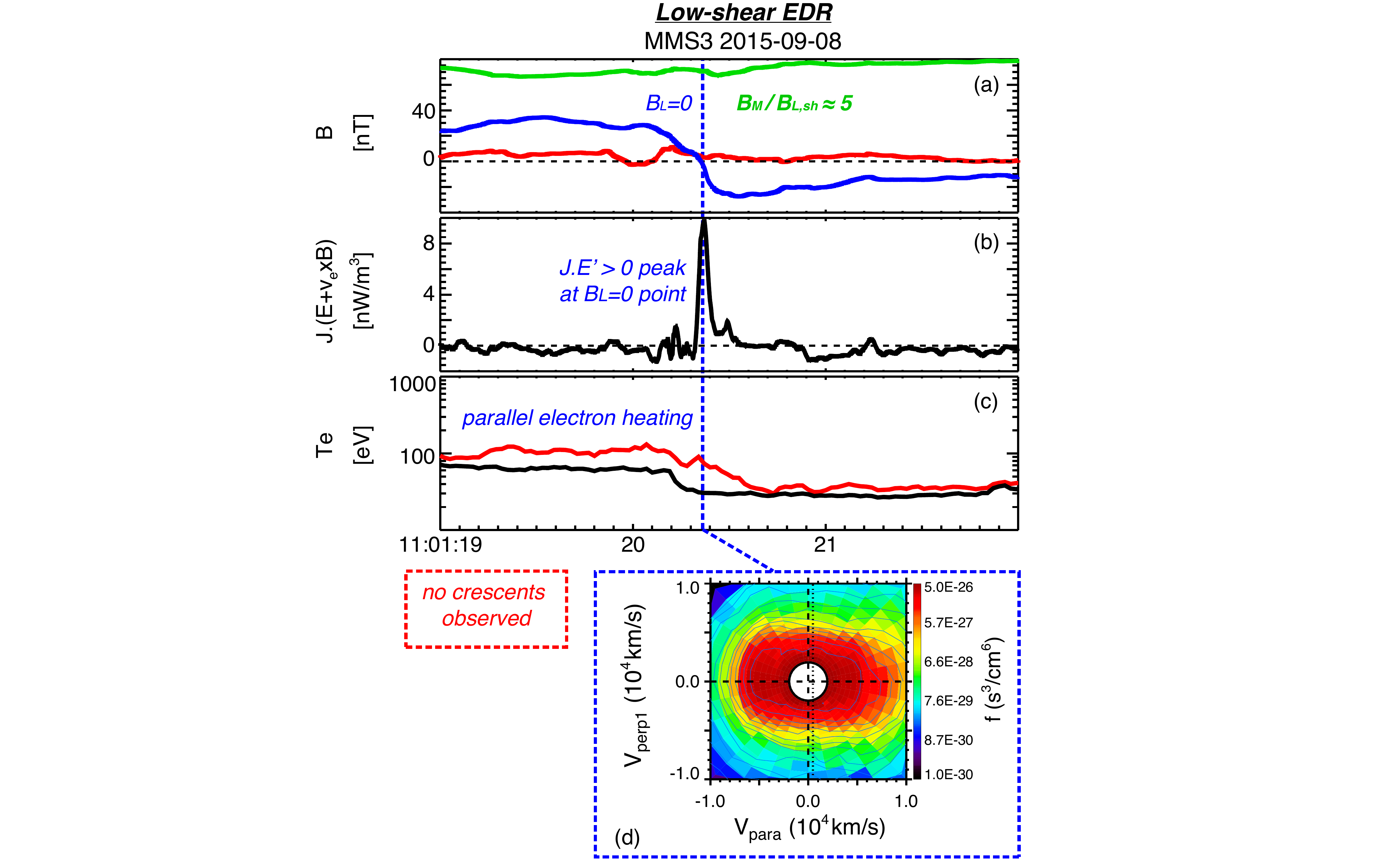}
\caption{Similar to Figures \ref{hsplots} and \ref{msplots} but for the very low-shear EDR of \citet{Eriksson.2016}, which also calculated the LMN coordinate system used here.}
\label{lsplots}
\end{figure*}

For the low-shear EDR event of \citet{Eriksson.2016}, which is shown in Figure \ref{lsplots}, no clear crescent-shaped eVDFs were observed, meaning that we cannot locate the $\vec{J}\cdot\vec{E}'>0$ peak relative to any magnetosphere-side landmark. However, $\vec{J}\cdot\vec{E}'$ was sharply positively peaked only at the $B_L=0$ point. No second peak or secondary structure was observed. This single sharp $\vec{J}\cdot\vec{E}'>0$ peak was seen by both of the two spacecraft that observed the 8 September EDR \citep{Eriksson.2016}. There was no significant agyrotropy, which is consistent with the lack of crescent-shaped eVDFs. \citet{Eriksson.2016} pointed out that the electron gyroradius was smaller than the magnetic scale size, which should prohibit crescent formation \citep{Hesse.2016}.

\subsection{``Intermediate''-shear events}

\begin{figure*}
\noindent\includegraphics[width=43pc]{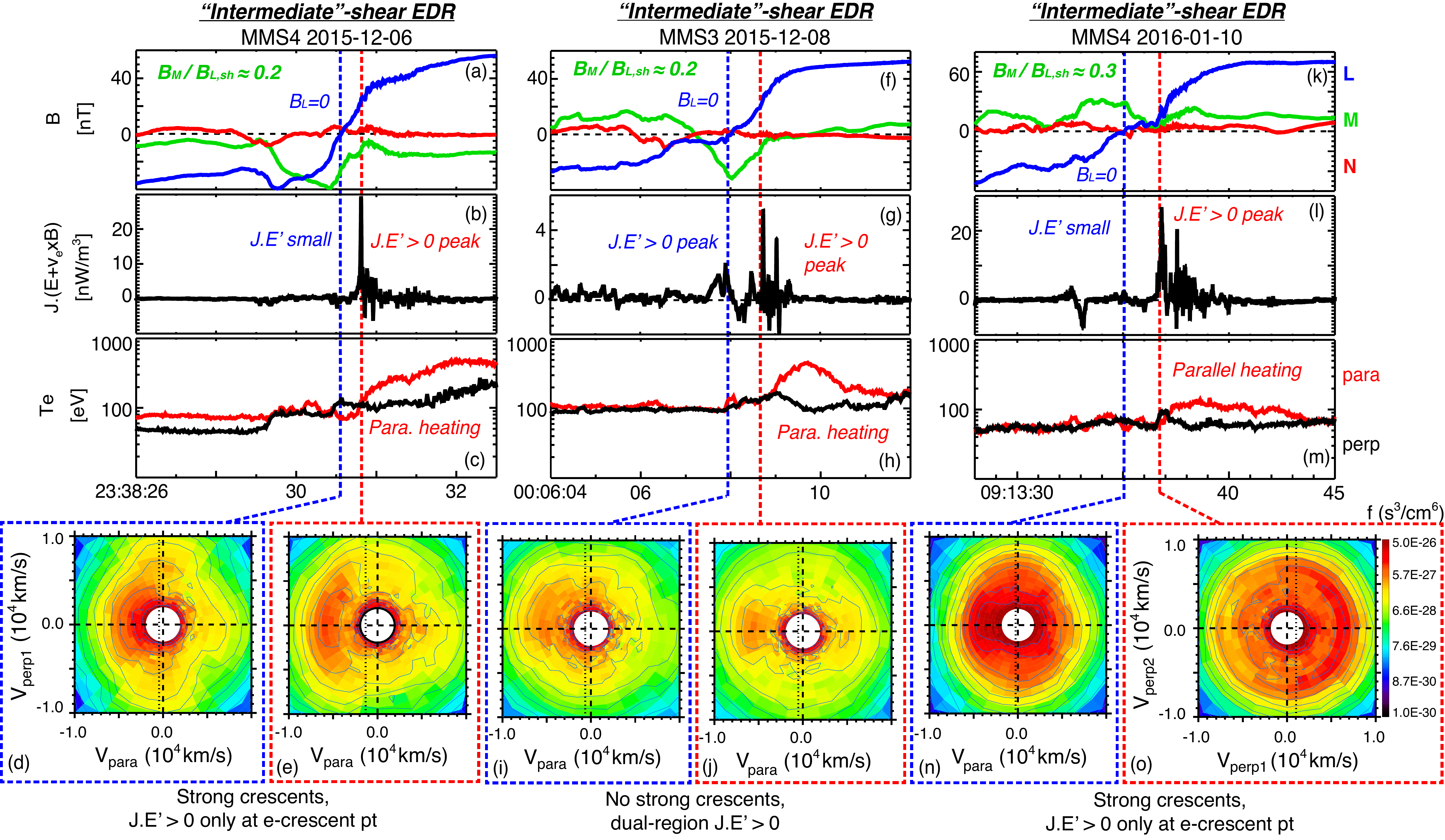}
\caption{(a, f, k) The magnetic field vector in LMN, (b, g, l) the energy conversion rate, $\vec{J}\cdot\vec{E}'$, and (c, h, m) the parallel (red) and perpendicular (black) electron temperatures for three ``intermediate''-shear EDR events with $B_M/B_{L,sh}\approx$20--30\%. The first column shows the event of \citet{Khotyaintsev.2016}, the second column is the event of \citet{Graham.2017}, and the third column \citet{Fuselier.2017}. The LMN systems for the events shown in the left-most and center columns were taken from the aforementioned studies. The LMN system for the third event was determined by applying MVAB to the data from MMS4 measured between 9:13:30 -- 9:13:42 UT. The axes are given by L = [0.107, --0.478, 0.872], M = [--0.809, --0.551, --0.203], and N = [0.578, --0.684, --0.446] in GSE X, Y, and Z.}
\label{isplots}
\end{figure*}

Figure \ref{isplots} shows data from the remaining three EDR events considered in this study, which we have categorized as having ``intermediate'' magnetic shear given that the events have $B_M/B_{L,sh}\sim20--30\%$, falling between our high ($B_M/B_{L,sh}\leq0.1$) and moderate-shear categories. Two of the three events exhibit essentially the same characteristics as the high-shear events, with $\vec{J}\cdot\vec{E}'>0$ and parallel electron heating only at the electron-crescent point and no activity at the $B_L=0$ point. The third intermediate-shear event, which occurred on 8 December 2015 ($\sim$11 hours before the 11:20 UT EDR event of \citet{BurchandPhan.2016}), had $\vec{J}\cdot\vec{E}'>0$ both at and Earthward of the $B_L=0$ point. For this third event, there is clear evidence of parallel heating Earthward of the $B_L=0$ point at the second and largest $\vec{J}\cdot\vec{E}'>0$ peak, but no significant anisotropy at or near the $B_L=0$ point. Again, the separation between the $B_L=0$ point and the Earthward-side $\vec{J}\cdot\vec{E}'>0$ peaks were well resolved for all three events.

\section{Summary and conclusions}

We analyzed the electron-frame energy conversion rate $\vec{J}\cdot(\vec{E}+\vec{v}_e\times\vec{B})\equiv\vec{J}\cdot\vec{E}'$ that occurred during the intermediate-shear ($B_M/B_{L,sh}\sim1$) 8 December 2015 EDR event of \citet{BurchandPhan.2016}. We found that the $\vec{J}\cdot\vec{E}'$ region was highly structured, with $\vec{J}\cdot\vec{E}'\approx$10 nW/m$^3$ near the field reversal point, lower amplitude and patchy $|\vec{J}\cdot\vec{E}'|\leq5$ nW/m$^3$ Earthward of the X-point, and strong $\vec{J}\cdot\vec{E}'\approx$--10 nW/m$^3$ near the Earthward edge of the $\vec{J}\cdot\vec{E}'$ region. The strong positive $\vec{J}\cdot\vec{E}'$ was associated with a current carried by a counter-streaming beam-like portion of the eVDF, which was partially balanced by a lower-energy parallel magnetosheath-inflow-like portion of the eVDF. The strong negative $\vec{J}\cdot\vec{E}'$ was associated with the turning of this beam into the $\hat{v}_{\bot1}$ direction.

We calculated $\vec{J}\cdot\vec{E}'$ for 10 other previously published EDR events with differing guide field strengths. For three nearly anti-parallel events, $\vec{J}\cdot\vec{E}'>0$ and parallel electron heating were only observed Earthward of the $B_L=0$ point at the electron-crescent point. Two of three ``intermediate''-shear EDRs had these same characteristics. For one of the three ``intermediate''-shear EDRs, as well as for three moderate-shear EDRs (not including the 8 December event of \citet{BurchandPhan.2016}), $\vec{J}\cdot\vec{E}'>0$ was observed at both the $B_L=0$ and electron-crescent points. For some but not all of these dual-region $\vec{J}\cdot\vec{E}'>0$ events, the intensity of the crescent-shaped portion of the eVDF was considerably reduced as compared to the anti-parallel events. Lastly, for one low-shear EDR, $\vec{J}\cdot\vec{E}'>0$ was only observed at the $B_L=0$ point and no crescent-shaped eVDFs were detected.

\subsection{Interpretation: influence of shear angle on energy conversion}

From this collection of cases it appears that the introduction of a guide field enhances $\vec{J}\cdot\vec{E}'$ at the X-point, which is similar to a result of \citet{PritchettandMozer.2009}, as discussed in the first section of this paper. The possible mechanism for this switching is easily explained. The introduction of a guide field causes the magnetic field at the X-point to be non-zero, allowing for free streaming of the electrons along the guide field due to the out-of-plane reconnection electric field. This is consistent with the eVDFs near the $B_L=0$ points from the intermediate-to-low-shear EDR events of 28 November 2016, 8 December 2015, and 8 September 2015, which had guide fields 0.5, 1, and 8 times as large as $B_{L,sh}$, respectively. For these events, the current near the X-point and at the $\vec{J}\cdot\vec{E}'$ peak was carried by electrons with magnetosheath-inflow-like VDFs shifted in the $+v_M$ direction. There was significant structure to the eVDF near the X-point for the 8 December (11:20 UT) event, which did not appear in the eVDFs for the other two aforementioned events. This structured eVDF may be a result of the larger electric field for this 8 December event, which was roughly 10 and 4.3 times the size of the parallel electric field for the 28 November and 8 September events, respectively. 

There is also an apparent trend, based on these few events, where increasing the guide field reduces $\vec{J}\cdot\vec{E}'$ near the electron-crescent point. In the 28 November event, with $B_M\sim0.5B_L$, strong parallel crescents were observed. In the 8 December (11:20 UT) event, with $B_M\sim B_L$, weaker parallel and perpendicular crescents were observed \citep{BurchandPhan.2016}. No crescents were observed for the 8 September event \citep{Eriksson.2016}. Increasing the strength of the guide field may reduce the intensity of the crescents, which are a result of finite gyroradius effects \citep{Hesse.2014,Hesse.2016}, by reducing the ratio of the gyroradius to the skin depth. 

It should be noted, however, that the upstream conditions were not uniform for these events. Table 1 lists some of the dimensionless parameters for asymmetric reconnection. As discussed in Section 1, the displacement of the flow stagnation point from the X-point depends on, among other factors, the density asymmetry $n_{sh}/n_{sph}$ and the magnetic field asymmetry $B_{L,sh}/B_{L,sph}$. The density asymmetries ranged from 2.5 to 40, but there is no obvious correlation with the location of the $\vec{J}\cdot\vec{E}'$ region with this parameter. For example, both the low-shear event of \citet{Eriksson.2016} ($\vec{J}\cdot\vec{E}'>0$ at X-point) and the moderate-shear event of \citet{BurchandPhan.2016} (dual-region $\vec{J}\cdot\vec{E}'>0$) had density asymmetries of 2.5, on the lowest end of this parameter range, and the locations of $\vec{J}\cdot\vec{E}'>0$ differed significantly. The 28 November event, which had a guide field of $\sim$0.5 and a density asymmetry of $30$, on the highest end of the parameter range, also had dual-region $\vec{J}\cdot\vec{E}'>0$ similar to the event of \citet{BurchandPhan.2016}. Similar comparisons can be made with the asymmetries of $B_L$, $T_e$, and $T_i$, where several events with similar parameter values can be found in different shear categories with different $\vec{J}\cdot\vec{E}'>0$ locations. Given that the locations of $\vec{J}\cdot\vec{E}'>0$ are well organized by the strength of the guide field (as compared to organization by any other single parameter), we suggest that the strength of the guide field plays a dominant role in controlling the location of $\vec{J}\cdot\vec{E}'>0$ for asymmetric reconnection.

The three ``intermediate''-shear events had similar guide field strengths of $B_M/B_{L,sh}\sim0.2$, yet only one of the three had dual-region $\vec{J}\cdot\vec{E}'>0$. The other two events in this category had $\vec{J}\cdot\vec{E}'>0$ only at the electron-crescent point, similar to the high-shear events. The differences between events in this category may indicate that there are other factors beyond the strength of the guide field that control the location of the $\vec{J}\cdot\vec{E}'>0$ region and/or that our determination of the upstream conditions was inexact. This approximate value of $B_M/B_{L,sh}\sim0.2$ may also be unique, as it is thought to be in this range that symmetric reconnection transitions between anti-parallel-like and component-like \citep{Swisdak.2005}.

\subsection{Future work}

This study was based on a small set of events. A more comprehensive analysis and characterization of EDR events with varying guide field strengths should be conducted to confirm or refute the interpretation provided in the previous section. Additional low-shear EDRs should be identified and/or included in this analysis. As of yet, to the knowledge of the authors, there has only been one very low-shear EDR event identified in the MMS data. This limited number of events may be explained if (a) the guide field suppresses crescent formation and (b) crescents are being used to identify EDRs. 

Another question is related to strength of the density asymmetry. The density asymmetries varied between these events and, as discussed previously, separation of the X-point and the stagnation points and crescent formation are both consequences of asymmetric reconnection. This question may be most easily addressed with simulations, where all of the parameters for a reconnection event may be pre-defined.

We have largely introduced the differences between $\vec{J}\cdot\vec{E}'$ for low and high-shear reconnection in a phenomenological manner, so there are many open questions related to the underlying physics that should create these differences. The mechanism for electron acceleration near the X-point during component reconnection and the parameters that govern the separation between $\vec{J}\cdot\vec{E}'$ peaks for intermediate-shear reconnection are both unknown. It is also unknown why, despite having similar upstream conditions, $\vec{J}\cdot\vec{E}'$ was an order of magnitude smaller for the 28 November event than for the 8 December and 9 September events. 


%
%
%
%
%
%
%

\begin{acknowledgments}
The authors would like to thank everyone who contributed to the success of the MMS mission and those who contributed to the rich scientific heritage on which this mission is based. This work was funded by the Austrian Academy of Sciences FFG project number 847069. Kevin Genestreti would like to thank Rumi Nakamura, Yasuhito Narita, Stephen Fuselier, and Jerry Goldstein for helpful conversations. The MMS data are publicly available at https://lasp.colorado.edu/mms/sdc/public except for the level 3 EDP data, which is currently only available via special request from the instrument team. 
\end{acknowledgments}

\end{article}
%
%
%
%
%
%
%
%


\end{document}